\documentclass[prb,twocolumn,amsmath,amssymb,]{revtex4}

\usepackage{graphicx}
\usepackage{dcolumn}
\usepackage{bm}

\newcommand{\lco}{LaCoO$_3$}
\newcommand{\lsco}{La$_{1-x}$Sr$_x$CoO$_3$}

\newcommand{\lmco}{La$_{1-x}$M$_x$CoO$_3$}
\newcommand{\lcco}{La$_{1-x}$Ca$_x$CoO$_3$}
\newcommand{\lbco}{La$_{1-x}$Ba$_x$CoO$_3$}
\newcommand{\lmmo}{La$_{1-x}$M$_x$MnO$_3$}

\newcommand{\cels}{$^\circ $C}

\newcommand{\etal}{{\it et~al.}}

\newcommand{\LS}{$t_{2g}^{\,6}e_{g}^0$}
\newcommand{\IS}{$t_{2g}^{\,5}e_{g}^1$}
\newcommand{\HS}{$t_{2g}^{\,4}e_{g}^2$}

\begin{document}

\title{Structure, Magnetization and Resistivity of \lmco\ (M~=~Ca, Sr, and Ba)}

\author{M.~Kriener}
\author{C.~Zobel}
\author{A.~Reichl}
\author{J.~Baier}
\author{M.~Cwik}
\author{K.~Berggold}
\author{H.~Kierspel}
\author{O.~Zabara}
\author{A.~Freimuth}
\author{T.~Lorenz}
\affiliation{ II.~Physikalisches Institut, Universit\"{a}t zu K\"{o}ln,
Z\"{u}lpicher Str. 77, 50937 K\"{o}ln, Germany}

\date{\today}

\begin{abstract}

We present an investigation of the influence of structural
distortions in charge-carrier doped \lmco\ by substituting
La$^{3+}$ with alkaline earth metals of strongly different ionic
sizes, that is M\,=\,Ca$^{2+}$, Sr$^{2+}$, and Ba$^{2+}$,
respectively. We find that both, the magnetic properties and the
resistivity change non-monotonously as a function of the ionic
size of M. Doping \lmco\ with M~=~Sr$^{2+}$ yields higher
transition temperatures to the ferromagnetically ordered states
and lower resistivities than doping with either Ca$^{2+}$ or
Ba$^{2+}$ having a smaller or larger ionic size than Sr$^{2+}$,
respectively. From this observation we conclude that the
different transition temperatures and resistivities of \lmco\ for
different M (of the same concentration $x$) do not only depend on
the varying chemical pressures. The local disorder due to the
different ionic sizes of La$^{3+}$ and M$^{2+}$ play an important
role, too.

\end{abstract}


\maketitle

\section{Introduction}

Among transition-metal oxides with perovskite structure ABO$_3$
the compound \lco\ is of particular interest because it shows a
spin-state transition as a function of temperature. In its ground
state \lco\ is a non-magnetic insulator, but with increasing
temperature a paramagnetic insulating state continuously develops
above about 50\,K and around 500\,K an insulator-to-metal
transition is observed. The spin-state transition may be
attributed to the fact that Co$^{3+}$ with 3d$^6$ configuration
can occur in different spin states depending on the ratio of
Hund's rule coupling and crystal field splitting. The ground
state of \lco\ is usually attributed to the low-spin
configuration of Co$^{3+}$ (LS: \LS ; S\,=\,0). However, the
question whether the paramagnetic behavior above 100\,K arises
from a thermal population of the high-spin state (HS: \HS ;
S\,=\,2) or of the intermediate-spin state (IS: \IS ; S\,=\,1) is
subject of controversial discussions. Earlier
publications~\cite{raccah67a,asai94a,itoh94a,senaris95a,yamaguchi96a}
often assume a population of the HS state whereas more recent
investigations~\cite{potze95a,korotin96a,saitoh97b,asai98a,yamaguchi97a,kobayashi00b,zobel02a}
often favor a LS/IS scenario.

Another aspect of cobaltates that is controversially discussed in
literature is the influence of charge carrier doping which can be
obtained by partial substitution of three-valent La$^{3+}$ by
divalent alkaline earth metals such as Ca$^{2+}$, Sr$^{2+}$, or
Ba$^{2+}$. Most studies concern the \lsco -series, where it is
found that the non-magnetic ground state is rapidly suppressed
with increasing $x$. For small $x$ a spin-glass behavior is
observed at low temperatures whereas larger doping leads to a
ferromagnetic order.\cite{itoh94a,sathe96a} The resistivity
strongly decreases with $x$ and above $x\sim 0.2$ metallic
behavior is observed. These qualitative features are always found
in studies on \lsco , but the detailed phase diagrams presented
so far are contradictory.\cite{itoh94a,senaris95b} In some samples
the transition temperatures to a spin glass or a ferromagnetic
state monotonously increase with increasing Sr
content,\cite{itoh94a} whereas other samples show anomalies
around 250\,K which occur almost independent of the Sr content
for $x\ge 0.1 $.\cite{caciuffo99a,senaris95b} It has been
proposed that these differences arise from different preparation
techniques.\cite{golovanov96a,anilkumar98a,joy99a} Moreover, it is
not clarified which spin states are present and whether or not
spin-state transitions take place as a function of temperature in
the Sr-doped systems.\cite{munakata97a,chainani92a,tsutsui99a}

The electronic and magnetic properties of doped \lco\ will not
depend on the charge carrier concentration alone but also on
structural parameters resulting e.\,g.\ from chemical pressure.
During the last decade the interplay between charge carrier
doping and chemical pressure has been intensively studied in
doped manganites and a broad variety of fascinating physical
phenomena has been observed (for recent reviews see
e.\,g.~[\onlinecite{salamon01a,rao00a,coey99a}]). For doped
cobaltates there are only a few studies on \lbco\
~[\onlinecite{patil79a,fauth01a,wang02b}] and \lcco
~[\onlinecite{taguchi82a,zock95a,yeh97a,samoilov98a,ganguly99a,muta02a,baily02a,baily03a}].
These data indicate qualitative similarities between Ca, Sr, and
Ba doping with respect to the suppression of the non-magnetic
ground state and to the occurrence of some kind of magnetic
order. In this paper we present a comparative study of the
structural data as well as magnetization and resistivity
measurements of the series \lcco , \lbco , and \lsco .

\section{Preparation and Structure}

Our samples were prepared by a standard solid-state reaction using
La$_2$O$_3$, Co$_3$O$_4$ and MCO$_3$ (M~=~Ca, Sr, Ba) as starting
materials. The materials were mixed in the prescribed ratio and
calcined in a Pt crucible at 850\,\cels\ for 48\,h in air. Then
the material was repeatedly reground and sintered at 1200\,\cels\
for 40\,h in air. For a bare polycrystal preparation the material
was finally pressed to cylindrical pellets (\O$\sim 13$\,mm;
$h\sim 1$\,mm) by 7.5\,kbar and sintered at 1200\,\cels\ for
40\,h in oxygen flow. For a single crystal growth the
polycrystalline material was pressed to a cylinder (\O$\sim
8$\,mm; $h\sim 10$\,cm) by 2\,kbar and sintered once more at
1200\cels\ for 20\,h in air. Single crystals of \lsco\ with $0\le
x \le 0.3$ were grown by the traveling zone method in a 4-mirror
image furnace (FZ-T-10000-H-VI-VP, Crystal Systems Inc.) under
oxygen atmosphere of 5\,bar with typical growth velocities of
3.5\,mm/h. All crystals are single phase as confirmed by powder
X-ray diffraction and the Sr content, determined by energy
dispersive X-ray diffraction (EDX), agrees within few percent to
the nominal composition. Single crystallinity was confirmed by
neutron diffraction and by Laue photographs. However, all
crystals are heavily twinned as it is usually the case for
slightly distorted perovskites.

In this way we also prepared single phase polycrystals of Ba- and
Ca-doped \lmco\ with $0\le x \le 0.3$. However, a single crystal
growth turned out to be much more difficult in these cases. From
\lbco\ we prepared a large single crystal with $x=0.1$, but a
growth starting with $x=0.2$ broke down after 2\,cm and the
obtained rod had a Ba content of only $x\simeq 0.13$. This
problem was even more severe for \lcco . Growth efforts starting
with $x=0.05$ and $0.1$ yielded crystals with Ca contents of only
$x\simeq 0.03$, whereas the EDX analysis of the remaining melts of
these efforts revealed Ca contents up to $x\simeq 0.4$. Therefore
we suspect solubility limits of Ca and Ba around $x\simeq 0.03$
and $\simeq 0.13$ for our growth conditions,\cite{remark_lcco_sc}
whereas significantly higher doping concentration can be reached
by a pure solid-state reaction. The measurements presented below
were performed on polycrystals of \lbco\ and \lcco\ whereas in
the case of \lsco\ single crystals have been investigated.

\begin{figure}
\includegraphics[angle=270,width=8cm,clip]{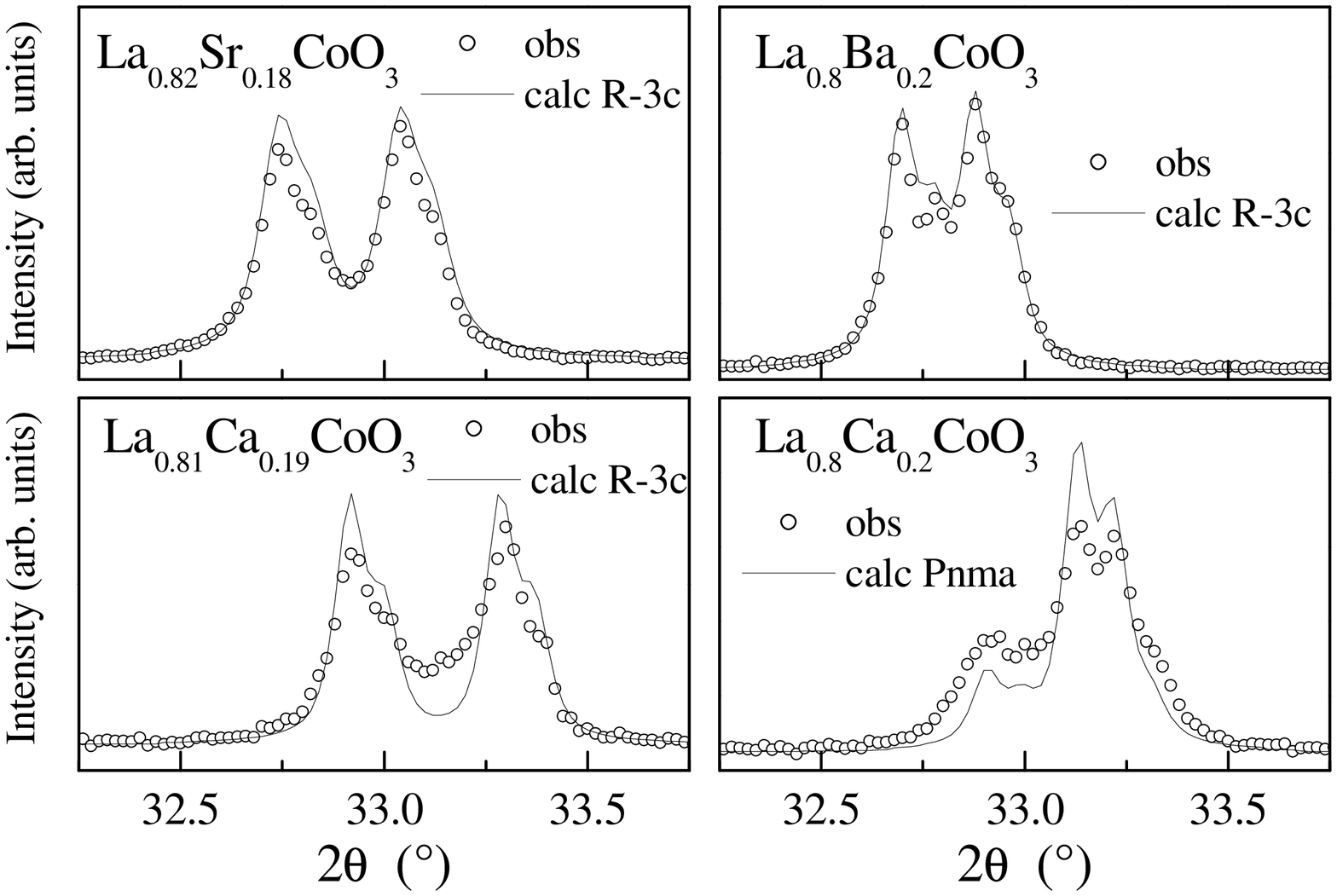}
 \caption[]{Representative X-ray diffraction patterns of \lmco\
 for $\rm M=Sr$, Ba, and Ca with $0.18\le x\le0.2$ around
 $2\theta=33^\circ$. The solid lines are calculated
 intensities. There is a structural change of \lcco\ from rhombohedral
 ($R\bar{3}c$ for $x\le 0.19$) to orthorhombic symmetry ($Pnma$ for $x\ge
 0.2$).} \label{xray}
\end{figure}

The diffraction patterns of \lsco\ and \lbco\ can be indexed by a
rhombohedral unit cell containing 2 formula units (space group
$R\bar{3}c$). We note that a recent high-resolution single
crystal X-ray study of the undoped \lco\ revealed a small
monoclinic distortion, which is related to a Jahn-Teller effect
of the thermally excited Co$^{3+}$ ions  in the intermediate-spin
state,\cite{maris03a} but the observed distortion in \lco\ is too
small to be observed in a usual powder X-ray analysis. \lcco\
with $x \le 0.19$ also has rhombohedral symmetry, but for $x\ge
0.2$ there is a structural change to orthorhombic symmetry
(Pnma), which can be best visualized around $2\theta= 33^\circ$
in the diffraction pattern. As shown in Fig.~\ref{xray} the
patterns of the Sr- and Ba-doped samples and for the Ca-doped
sample with $x=0.19$ are very similar and well described by
$R\bar{3}c$ symmetry. However, the pattern of \lcco\ with $x=0.2$
is systematically different and a description within $Pnma$
symmetry is clearly better than within $R\bar{3}c$. In this
respect our result differs from those of
Refs.\,[\onlinecite{taguchi82a,zock95a}], where the rhombohedral
symmetry was used also for $x>0.2$, and of
Ref.\,\onlinecite{muta02a}, where a two-phase region consisting
of an admixture of a rhombohedral and a pseudo-cubic phase was
proposed for $x>0.2$.

\begin{figure}
\includegraphics[angle=0,width=8cm,clip]{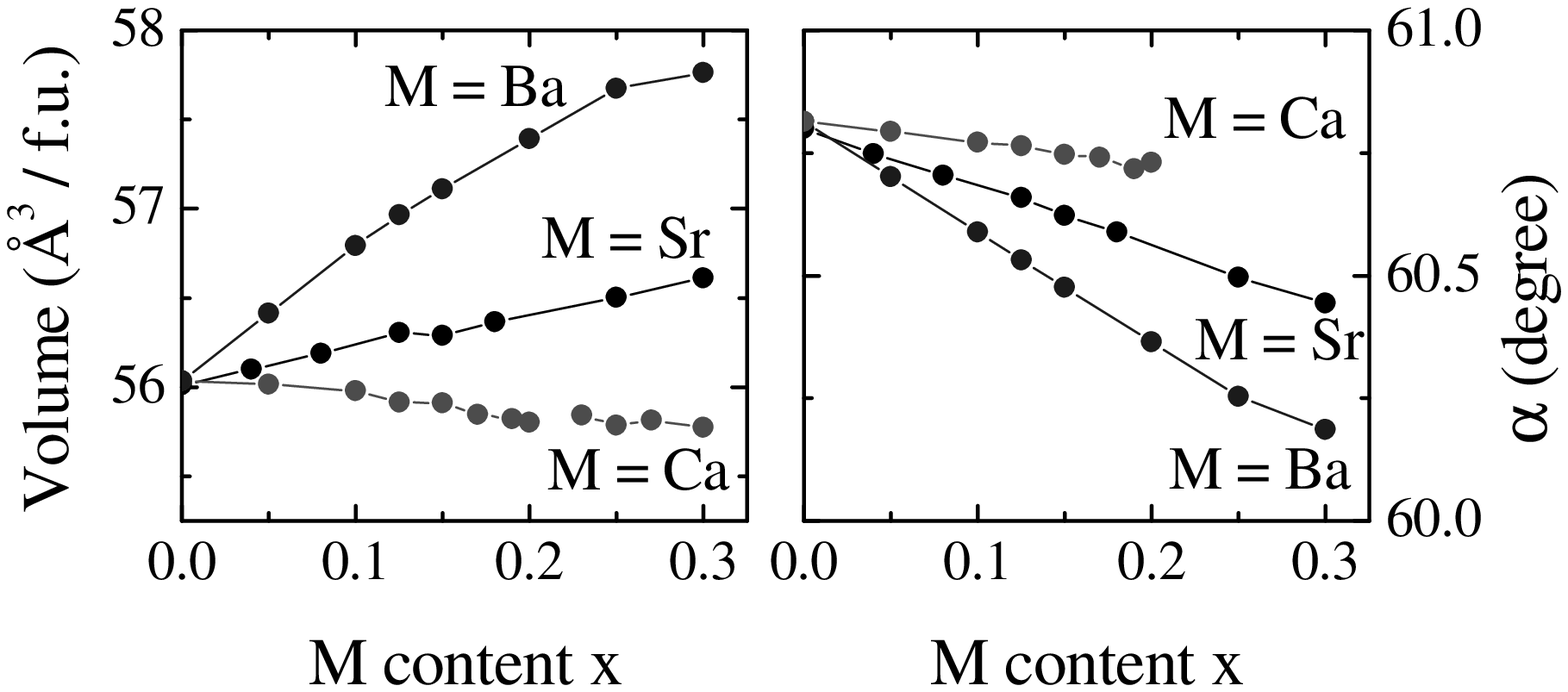}
 \caption[]{Volume per formula unit (left) and rhombohedral
 angle $\alpha_{\rm R}$ (right) as a function of doping for
 \lmco\ with $\rm M=Ba$, Sr, and Ca. In \lcco\
 symmetry changes to orthorhombic ($Pnma$) for $x\ge
 0.2$.} \label{cellvol}
\end{figure}

In Fig.~\ref{cellvol} we compare the doping dependencies of the
volume per formula unit (f.u.) and of the rhombohedral angle
$\alpha_{\rm R}$ for all three series. The volume increases
strongly with increasing Ba and moderately with increasing Sr
content whereas Ca doping causes a slight decrease of the unit
volume. For all three dopings the rhombohedral angle
systematically decreases with increasing content of M. For Ca
doping a transition to an orthorhombic lattice takes place,
whereas for Ba and Sr cubic symmetry ($\alpha_{\rm R} =
60^\circ$) is approached and from a linear extrapolation of our
data is expected to occur around $x=0.37$ and 0.67, respectively.
This value agrees well with the observed cubic symmetry in \lbco\
for $x=0.4$,\cite{patil81a} but is too large for \lsco\ where a
cubic structure was reported already for $x=0.5$.\cite{bhide75a}

\section{Magnetization}

\begin{figure}
\includegraphics[width=8cm,clip]{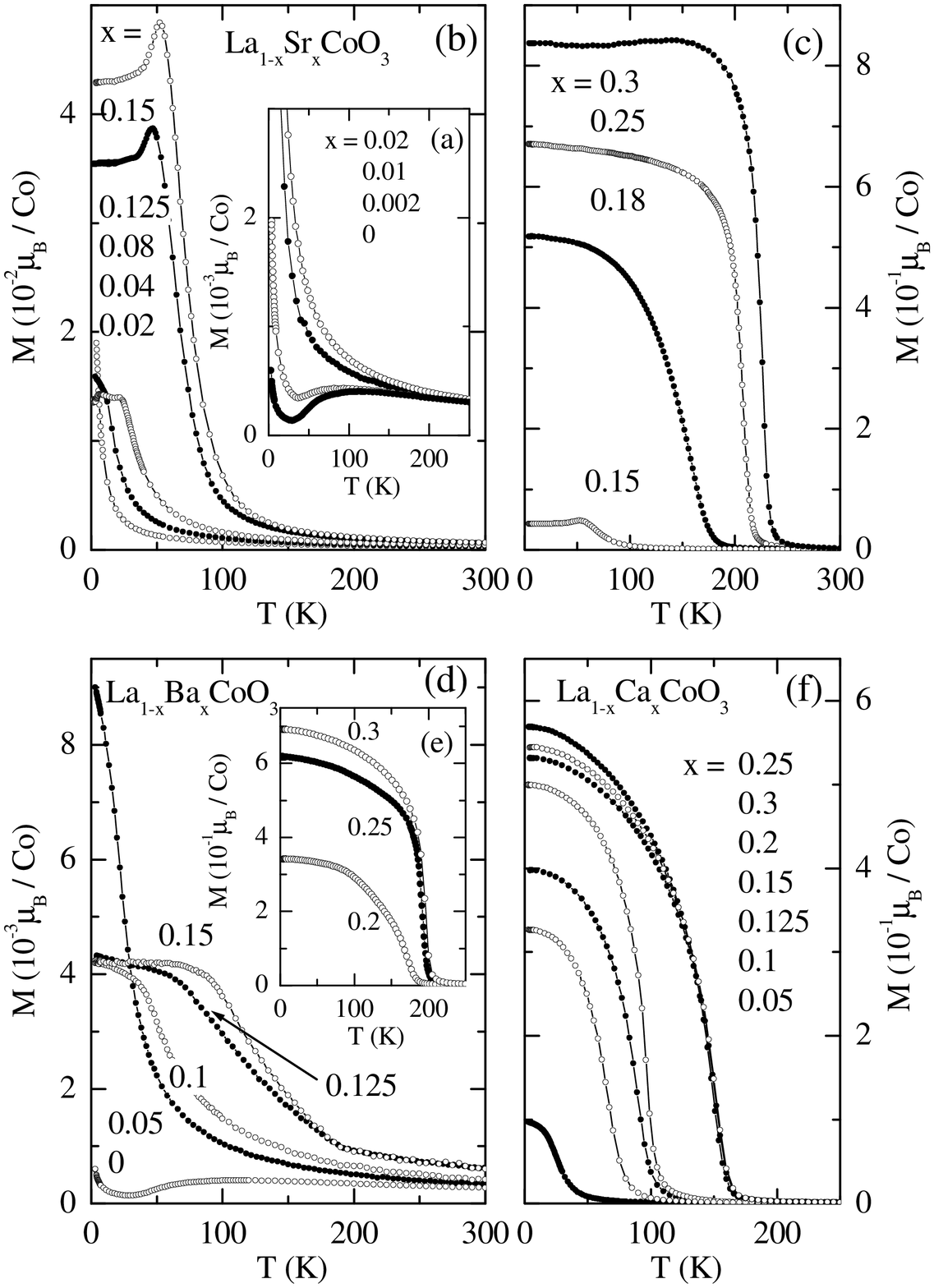}
 \caption[]{Magnetization vs.\ temperature of \lsco\ (a, b, c), \lbco\
 (d,e), and \lcco\ (f). All curves were taken during field-cooled
 runs in an applied magnetic field of 50\,mT. Please note the different orders of
 magnitude of $M$ for the different panels.} \label{lcochi}
\end{figure}

In Fig.~\ref{lcochi} we compare the magnetization of all three
\lmco\ series as a function of temperature. A magnetic field of
50\,mT has been applied at 300\,K, then the sample was cooled to
4\,K and the data were taken during the subsequent heating run.
The undoped \lco\ shows a pronounced maximum of $M(T)$ around
90\,K arising from the spin-state transition of the Co$^{3+}$ ions
(Fig.~\ref{lcochi}a). In Ref.~[\onlinecite{zobel02a}] we have
presented a combined analysis of this $M(T)$ curve and the
thermal expansion of \lco , which gives evidence for a low-spin
to intermediate-spin state scenario. A maximum of $M(T)$ is also
present for $x=0.002$, but not visible anymore for $x\ge 0.01$.
The samples with $0.04 \le x\le 0.15$ show a kink or a peak in
$M(T)$ indicating a spin-glass behavior at low temperatures
(Fig.~\ref{lcochi}b), whereas the crystals with $x\ge 0.18$ show
a continuously increasing spontaneous magnetization with
decreasing temperature as expected for a usual ferromagnet
(Fig.~\ref{lcochi}c). At low temperatures all \lsco\ crystals
show differences between the $M(T)$ curves obtained in the FC run
and the corresponding curves obtained in a ZFC run (not shown),
when the magnetic field is applied at the lowest temperature
after the sample has been cooled in zero field. Such a difference
is expected for a spin glass, but this difference alone does not
prove spin-glass behavior, since it can also arise from the
domain structure of a ferromagnet. In order to distinguish
between spin glass freezing or ferromagnetic order, one has to
perform time dependent studies, such as AC-susceptibility
measurements of different frequencies or relaxation studies of the
magnetization. From such studies it has been concluded that
\lsco\ shows a spin-glass behavior for $x < 0.18$ and the
occurrence of a so-called cluster spin glass has been proposed
for $0.18 \le x \le 0.5$.\cite{itoh94a} Our results of $M(T)$
agree with such an interpretation, but for the reasons explained
above we cannot confirm with ambiguity this conclusion. For the
sake of simplicity we will not try to discriminate between a
cluster spin glass and a ferromagnet in the following.

\begin{figure}
\begin{center}\includegraphics[width=8cm,clip]{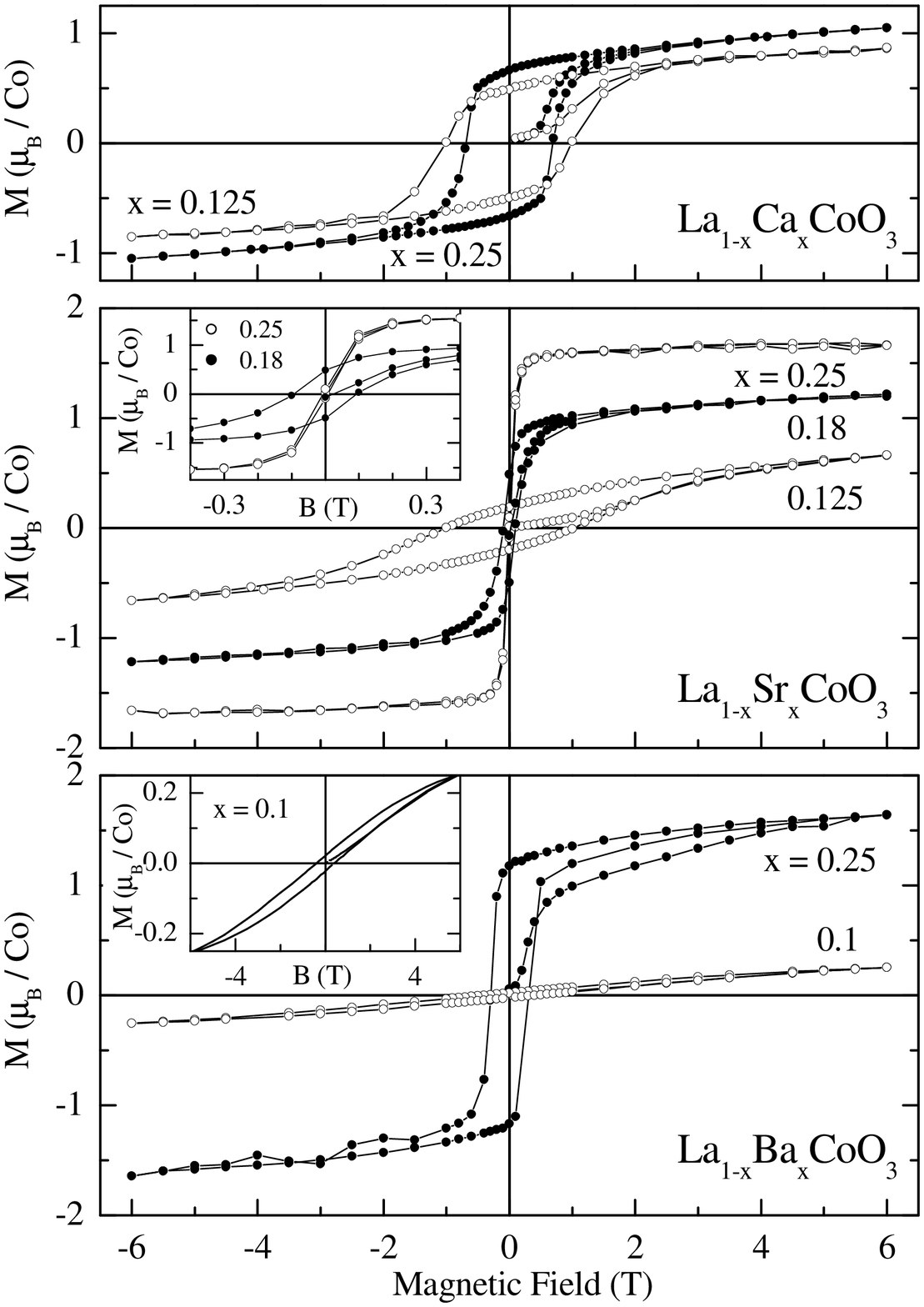}\end{center}
 \caption[]{Magnetization curves of \lcco\ (top), \lsco\ (middle),
 and \lbco\ (bottom) measured at $T=4$\,K. The inset of the middle panel is
 an expanded of the low-field region for the Sr-doped samples with $x=0.18$
 and $0.25$ showing that the hysteresis of the latter sample is extremely
 small. The inset of the bottom panel is an expanded view of $M(B)$
 of La$_{0.9}$Ba$_{0.1}$CoO$_{3}$ which has a small hysteresis over the entire
 field range.} \label{lmcomag}
\end{figure}

The magnetization of the Ba-doped samples with $x=0.05$ and~0.1
continuously increases with decreasing temperature and below a
certain temperature the curves flatten as is clearly seen for the
$x=0.1$ sample (Fig.~\ref{lcochi}d). Although more difficult to
see, the flattening is also present  for $x=0.05$ as we have
identified from a plot of $1/M$ vs.\ $T$ (not shown). Insofar the
Ba-doped samples with $x=0.05$ and $x=0.1$ show a very similar
behavior to the Sr-doped ones with $x=0.04$ and 0.08. With
increasing Ba doping the flattening of $M(T)$ becomes more
pronounced and another clear kink in $M(T)$ occurs around 200\,K.
It is remarkable that all Ba-doped samples with $0.1\le x\le
0.15$ show almost the same low-temperature value of $M$. This is
in contrast to the corresponding Sr-doped samples, where $M$ at
the lowest temperatures continuously increases with $x$. In
addition, the Ba-doped samples do not show a peak in the FC
$M(T)$ curves. For higher Ba concentrations ($x\ge 0.2$) a
pronounced spontaneous magnetization occurs as in a conventional
ferromagnet and both, the absolute value of $M$ and the
transition temperature $T_c$ systematically increase with $x$
(Fig.~\ref{lcochi}e).

The $M(T)$ curves of \lcco\ (Fig.~\ref{lcochi}f) differ from those
of the Ba- and Sr-doped series because the principal temperature
dependence does hardly change as a function of Ca content.
Already the sample with $x=0.05$ shows a spontaneous
magnetization with a temperature dependence typical for a
ferromagnet and the absolute value of $M(4$\,K) is about one order
of magnitude larger than those of the corresponding Sr- and
Ba-doped samples. With increasing Ca concentration both, $T_c$ and
the absolute value of $M(4$\,K) systematically increase with $x$
up to $x=0.2$ and remain essentially constant for $0.2\le x \le
0.3$. The different $M(4$\,K) values which depend
non-monotonously on $x$ cannot be taken seriously, since for
technical reasons we had to use extremely small samples.
Therefore the error in weighing the samples is of order 10\,\%,
which is larger than the differences in $M(4$\,K).

Fig.~\ref{lmcomag} gives an overview of the magnetization curves
measured at 4\,K for all three series. \lcco\ shows the typical
behavior of a ferromagnet with a large hysteresis region. With
increasing $x$ the hysteresis narrows and the absolute value of
the magnetization increases slightly. For both samples the
magnetization strongly flattens in the higher field range but the
saturation is apparently not yet reached. For the \lsco\ series
the form of the magnetization curves changes as a function of
$x$. The $x=0.25$ sample has only a very small hysteresis (see
Inset of Fig.~\ref{lmcomag}) and the magnetization is essentially
saturated in a field of 6\,T. For $x=0.18$ the magnetization
becomes smaller and the hysteresis larger, but the $M(B)$ curve
still has the form of a typical ferromagnet. The form of $M(B)$
changes when the Sr content is reduced to $x=0.125$. The
hysteresis region becomes very broad and there is no steep change
of $M$ around the coercive field. This difference can be
interpreted as further indication of a transition from spin-glass
behavior  to a ferromagnet (or a cluster spin glass) around
$x\simeq 0.18$. A very broad hysteresis is also present in \lbco
. As shown in the lower Inset of Fig.~\ref{lmcomag} the Ba-doped
sample with $x=0.1$ has a small hysteresis that extends even up
to the maximum field of 6\,T. Similar $M(B)$ curves are found in
the entire range $0.05 \le x\le 0.15$ of Ba doping. The samples
with $x\ge 0.2$ have again more conventional $M(B)$ curves with
hysteresis widths which are somewhat smaller and larger than
those of the corresponding Ca and Sr-doped samples, respectively.
However, a closer inspection of Fig.~\ref{lmcomag} reveals
another anomaly: La$_{0.75}$Ba$_{0.25}$CoO$_3$ has a virgin curve
of $M(B)$ that is located {\em outside} of the subsequent
hysteresis loop over a large field range. This anomalous behavior
is well reproducible and occurs also for $x=0.3$. To the best of
our knowledge this anomalous virgin curves are neither expected
for a ferromagnet nor for a spin glass.

\section{Resistivity}

\begin{figure}
\begin{center}\includegraphics[angle=0,width=8cm,clip]{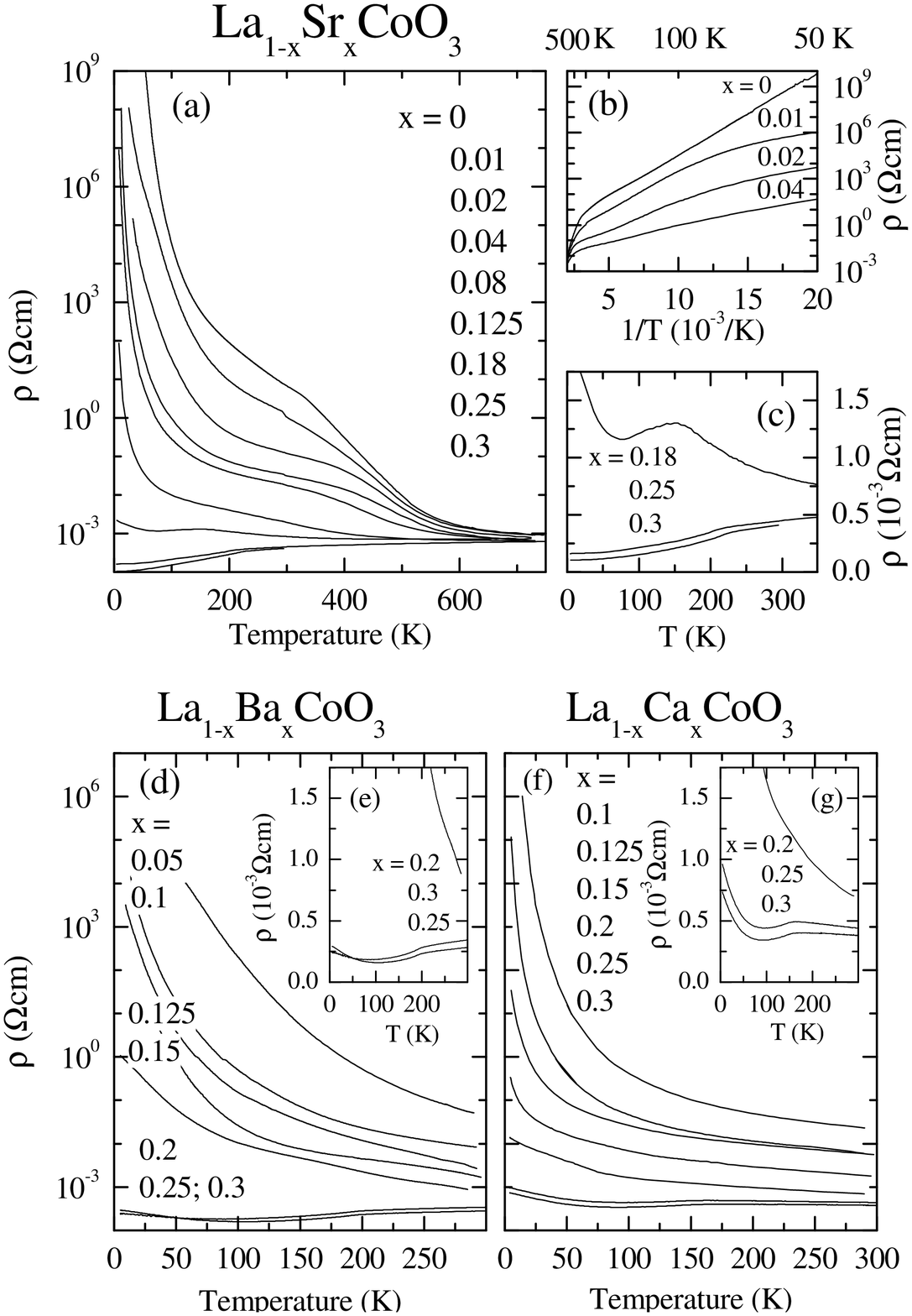}\end{center}
 \caption[]{Resistivity vs.\ temperature of \lsco\ (a,b,c), \lbco\ (d,e),
 and \lcco\ (f,g). Panel~(b) shows Arrhenius plots
 $\log(\rho)$ vs.\ $1/T$ for the low-Sr-doped samples ($x\le 0.04$) and
 panels~(c), (e), and (f) give expanded views of the
 low-temperature resistivities on a linear scale
 of the (nearly) metallic samples ($x\ge 0.18$).} \label{lcoR}
\end{figure}

Fig.~\ref{lcoR} compares the resistivity data of the \lmco\
series. In agreement with previous studies the undoped \lco\ is a
good insulator at low temperatures and shows an insulator-to-metal
transition around 500\,K. Below 400\,K $\rho$ shows an activation
type behavior $\rho \propto \exp(\Delta_{act}/T)$ down to 50\,K
with activation energy $\Delta_{act}\simeq 1240$\,K. The weakly
doped samples ($x\le 0.04$) show qualitatively similar
resistivity behavior as the pure compound, but the $\log(\rho)$
vs.\ $1/T$ curves are not linear (Fig.~\ref{lcoR}b). With further
increasing Sr content the low-temperature $\rho$ rapidly drops
and the samples above $x=0.18$ show a metallic temperature
dependence of $\rho$ over the entire temperature range
(Fig.~\ref{lcoR}a). The crystals with $x=0.18$, 0.25, and 0.3 show
distinct anomalies in $\rho(T)$ at the critical temperatures of
ferromagnetic order (Fig.~\ref{lcoR}c).

A drastic decrease of $\rho$ with increasing concentration is also
found for Ca and Ba doping, but there are no Ca- or Ba-doped
samples showing metallic resistivity behavior over the entire
temperature range. From the Ba-doped series only the samples with
$x\ge 0.25$ have a decreasing $\rho$ with decreasing temperature
from 300\,K down to about 100\,K with clear slope changes around
200\,K (Fig.~\ref{lcoR}e), where ferromagnetic order sets in.
Insofar the $\rho(T)$ curves of the higher Ba-doped samples are
similar to those of higher Sr-doped crystals, but with further
decreasing temperature the resistivities of the Ba-doped samples
increase again. From the Ca-doped series even the samples with
$x=0.25$ and $0.3$ show a weakly increasing $\rho$ with
decreasing temperature from 300\,K down to about 150\,K
(Fig.~\ref{lcoR}g). Then there is a slight decrease of $\rho(T)$
below about 150\,K, which is again the ferromagnetic ordering
temperature. Finally, both Ca-doped samples show a
low-temperature increase of $\rho$ below about 80\,K that is even
more pronounced than in the Ba-doped samples. One might suspect
that the low-temperature increase of $\rho$ arises from the fact
that the Ca- and Ba-doped samples are polycrystals whereas the
Sr-doped ones are single crystals. However, this is not the only
reason for the different low-temperature behavior. We have
measured $\rho$ of poly-crystalline La$_{0.7}$Sr$_{0.3}$CoO$_3$
and found that it also shows a low-temperature increase, but this
increase is weaker than in the Ba-doped and much weaker than in
the Ca-doped polycrystal.

\section{Discussion}

\begin{figure}
\begin{center}\includegraphics[width=8cm,clip]{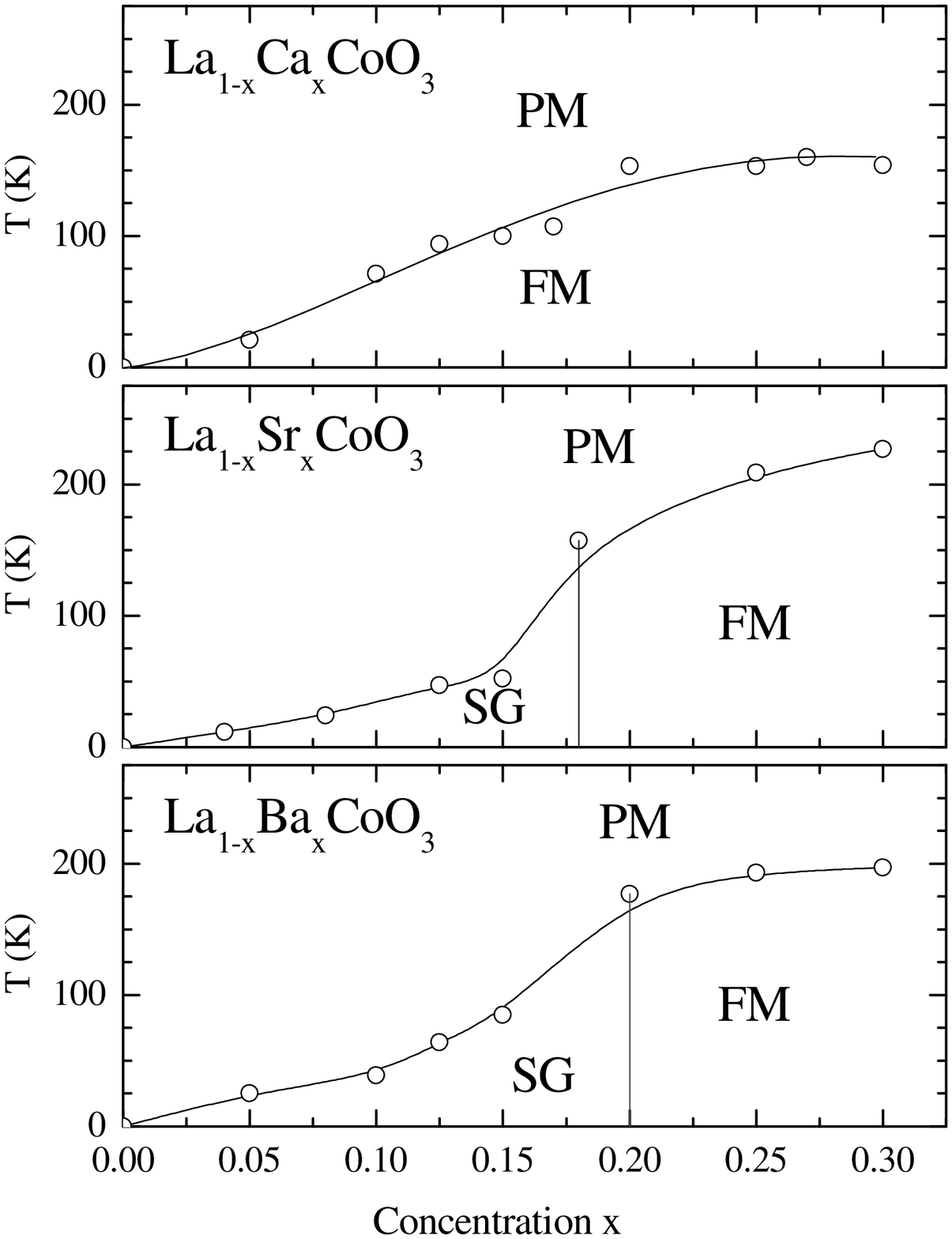}\end{center}
 \caption[]{Phase diagrams of \lmco\ with M=Ba, Sr, and Ca (from bottom to
 top). PM, FM, and SG stand for paramagnet, ferromagnet, and spin
 glass, respectively. The solid lines are to guide the eye. We mention, however, that
 an unambiguous distinction between
 a SG and a FM (or the so-called Cluster Glass Phase proposed in
 Ref.~\onlinecite{itoh94a} for \lsco\ with  $x > 0.18$) is not possible from our
 measurements of the static magnetization alone.} \label{phadis}
\end{figure}

In Fig.~\ref{phadis} we summarize the phase diagrams obtained from
the magnetization measurements. For \lcco\ we find transitions
from a paramagnetic to a ferromagnetic phase for the entire
concentration range.  The transition temperature $T_c$
monotonously increases with $x$ for $x\le 0.2$ and tends to
saturate around $T_c \sim 150$\,K for larger $x$. Our results of
the Sr-doped series nicely agree with those obtained by
Itoh~\etal.\cite{itoh94a} For $x\le 0.18$ the magnetization has
features which are typical for the proposed spin-glass behavior
and the characteristic temperature $T_c$ strongly increases when
$x=0.18$ is approached. For larger $x$ the $M(T)$ curves resemble
those of typical ferromagnets and $T_c$ increases up to about
230\,K for $x=0.3$. We emphasize that additional anomalous
features in the $M(T)$ curves around 240\,K, which have been
frequently observed in polycrystalline \lsco\ for a large doping
range $0.025\le x \le 0.5$,\cite{senaris95b,caciuffo99a} do not
occur in our single crystals. Thus we conclude that these
features are not intrinsic properties of \lsco\ but rather arise
from the different preparation technique of those polycrystals as
has been proposed in
Refs.\,[\onlinecite{golovanov96a,anilkumar98a,joy99a}]. The
Ba-doped series has qualitatively similar $M(T)$ curves as the
Sr-doped samples. Thus, we conclude that \lbco\ shows a
spin-glass behavior for $x<0.2$ and a ferromagnetic order for
larger $x$ with a saturation of $T_c$ around 200\,K.

For $x \ge 0.2$ all three series of \lmco\ show ferromagnetic
order and for a given $x$ the Sr-doped samples have the largest
$T_c$ values ($\sim 220$\,K), the Ba-doped samples have somewhat
smaller ($\sim 200$\,K) and the Ca-doped samples significantly
smaller $T_c$'s ($\sim 150$\,K). Obviously, $T_c$ depends
non-monotonously on the ionic size of M. Since the less
conducting \lcco\ samples have the lowest transition temperatures
one may suspect a correlation between metallic conductivity and
ferromagnetic order as it would be expected within a
double-exchange model. However, the Ba-doped samples have lower
$T_c$ values than the Sr-doped ones although their resistivities
are comparable or even slightly smaller than those of the
Sr-doped samples.

In doped manganites \lmmo\ $T_c$ also dependends non-monotonously
on the ionic sizes of  M = Ca, Sr, and Ba for $x\ge
0.15$.\cite{mandal03a,hwang95a,dabrowski98a} In the manganite
case it is argued that the transition temperature is mainly
determined by two kinds of structural
distortions.\cite{rodriguezmartinez96a} The first is a global
distortion arising from the deviation of the structure from the
cubic perovskite. This distortion can be described by the
deviation of the tolerance factor $t=(\langle
r_A\rangle+r_O)/(\sqrt{2}(r_{Mn}+r_O))$ from $t=1$ where $r_O$ and
$r_{Mn}$ denote the ionic radii of the Mn and O ions and $\langle
r_A\rangle=(1-x)\,r_{La}+x\,r_{M}$ is the average radius of the
ions on the A site, that is the average of the radii of La$^{3+}$
and M$^{2+}$\,=\,Ca$^{2+}$, Sr$^{2+}$, or Ba$^{2+}$,
respectively. The second type is a local distortion arising from
the different ionic radii of La$^{3+}$ and M$^{2+}$, that can be
described by the variance of the A-site ionic radii
$\sigma^2=(1-x)\,r_{La}^2+x\,r_{M}^2-\langle r_A\rangle^2$. In
order to quantify these two phenomena the following empirical
relation was proposed:\cite{rodriguezmartinez96a}
\begin{equation}
T_c(\langle r_A\rangle,\sigma)=T_c(r_A^0,0)-p_1Q_0^2-p_2\sigma^2
\,\, .
\end{equation}
Here, $T_c(\langle r_A\rangle,\sigma)$ is the real transition
temperature and $T_c(r_A^0,0)$ is a hypothetical transition
temperature of an ideal cubic perovskite with A-site ion radius
$r_A^0$ that fulfills $t=1$ and $\sigma=0$. The deviation from
cubic symmetry is given by $Q_0=r_A^0-\langle r_A\rangle$ and
$p_1$ and $p_2$ are constants. We use the tabulated ionic radii
$1.35$\,\AA\ for O$^{2-}$ and 1.216\,\AA, 1.18\,\AA, 1.31\,\AA,
and 1.47\,\AA\ for the A-site ions La$^{3+}$, Ca$^{2+}$,
Sr$^{2+}$, and Ba$^{2+}$ in 9-fold
coordination~\cite{shannon76a,remark12fold} and for
$r_A^0=0.6\,r_{\rm La}+0.4\,r_{\rm Ba} \simeq 1.32$\,\AA\ since
\lbco\ becomes cubic for $x=0.4$.\cite{patil79a} These values
yield for the remaining parameters $p_1\simeq 6.1\cdot
10^{3}$\,K/\AA$^2$, $p_2\simeq 3.4\cdot 10^{3}$\,K/\AA$^2$, and
$T_c(r_A^0,0) \simeq 229$\,K for $x=0.2$.\cite{remarkitoh} The
corresponding values for $x=0.25$ are $10.3\cdot
10^{3}$\,K/\AA$^2$, $6.3\cdot 10^{3}$\,K/\AA$^2$, and $286$\,K
and those for $x=0.3$ are $11.4\cdot 10^{3}$\,K/\AA$^2$,
$7.4\cdot 10^{3}$\,K/\AA$^2$, and $306$\,K.

For all three series the observed $T_c$ values are significantly
reduced from those of the corresponding hypothetical, ideal cubic
perovskites. For \lcco\ the reduction is most pronounced although
in this case the effect of the A-site disorder is negligibly
small ($\sigma^2$ is of order $10^{-4}$\,\AA$^2$) because the
ionic radii of La$^{3+}$ and Ca$^{2+}$ are very close to each
other. The reduction is almost completely a consequence of the
deviation from cubic symmetry which is most pronounced in \lcco .
This global distortion is already rather large in the undoped
\lco\ and increases with increasing substitution of La by the
smaller Ca ion. This is consistent with our observation of a
structural phase transition from rhombohedral to orthorhombic
which is expected for perovskites when the tolerance factor
becomes smaller. For Sr doping the A-site disorder is larger
($\sigma^2$ is of order $10^{-3}$\,\AA$^2$) than for Ca-doping,
but the $T_c$ reduction due to this effect remains small (of
order 10\,K). The Sr-doped samples have significantly larger $T_c$
values than the Ca-doped ones, since the structure approaches
cubic symmetry and therefore the reduction of $T_c$ due to the
global distortion is less pronounced. For Ba doping the structure
further approaches cubic symmetry and consequently the influence
of the global distortion on $T_c$ is further reduced. However,
the A-site disorder is now pretty large ($\sigma^2$ is of order
$10^{-2}$\,\AA$^2$), since Ba$^{2+}$ is much larger than
La$^{3+}$. In the case of Ba doping the $T_c$ reduction due to
the deviation from cubic symmetry is of only order $15$\,K, but
the reduction due to A-site disorder increases from about 30\,K
to 95\,K when $x$ increases from 0.2 to 0.3 and explains why the
$T_c$ values of \lbco\ do not exceed those of \lsco .

Considering the global and local distortions in \lmco\ also allows
to give some qualitative arguments for the different resistivity
behavior for different M. Since for the Ca-doped samples the
deviation from cubic symmetry is most pronounced, there is a
rather strong deviation of the Co-O-Co bond angle from
180$^\circ$ and the hopping integral, or in other words the
bandwidth, is small. As a consequence the $\rho(T)$ curves of
\lcco\ remain almost semi-conducting even for $x=0.3$. In \lsco\
the Co-O-Co angle is closer to 180$^\circ$ and consequently the
bandwidth increases leading to metallic $\rho(T)$ curves for $x>
0.18$. In \lbco\ the bandwidth will be even larger than in the
Sr-doped samples in agreement with the smaller resistivities we
observe at temperatures above about 100\,K. Yet, the $\rho(T)$
curves for Ba doping with $x\ge 0.25$ show a low-temperature
increase. This may arise from a larger local disorder in \lbco\
making these samples more sensitive to charge localization at low
temperatures than the Sr-doped samples.

Next we will discuss the doping range below $x=0.2$, where the
Sr- and Ba-doped series show spin-glass behavior which is absent
in \lcco . One source for spin-glass behavior are competing
ferromagnetic and antiferromagnetic interactions. Such a
competition may arise from local disorder which will be more
pronounced in the Sr- and Ba-doped samples due to the larger
A-site disorder than for Ca doping. Thus the absence of spin-glass
behavior in \lcco\ may arise from the similar ionic radii of
La$^{3+}$ and Ca$^{2+}$. Another reason could arise from the
structural phase transition of \lcco . This transition is located
at room temperature for $x=0.2$ and possibly shifts towards lower
temperatures for smaller $x$. If one assumes that the
ferromagnetic exchange is enhanced in the orthorhombic structure,
a further reason for the ferromagnetic order in low-doped \lcco\
could be that these samples are orthorhombic at low temperatures.

As mentioned in the introduction it is not clear which spin
states of the Co$^{3+}$ and Co$^{4+}$ ions are realized in \lmco\
samples, and it is not known whether there are spin-state
transitions as a function of temperature. From our present data
we cannot unambiguously resolve these puzzles. Clear indications
of spin-state transitions are only present in the $M(T)$ curves of
the \lsco\ samples with very low doping ($x< 0.01$). For larger
dopings the $M(T)$ curves are dominated by the occurrence of
spin-glass or ferromagnetically ordered phases. The saturation
values of the magnetization in the ferromagnetic phases with
$x=0.25$ amount to $M_S \sim 1 \mu_B/$Co for Ca- and to $\sim
1.65 \mu_B/$Co for Sr- and Ba-doping. The latter value fits best
to the expected spin-only value of $\sim 1.75 \mu_B/$Co expected
for a Co$^{4+}$ LS state ($t_{2g}^{\,5}e_{g}^0$) and a Co$^{3+}$
IS state ($t_{2g}^{\,5}e_{g}^1$). If these spin states are
realized one can easily imagine the relevance of the double
exchange mechanism and understand the similarity to the
manganites, because these spin states differ only by two
additional down-spin electrons in the $t_{2g}$ level from the
corresponding high-spin states of Mn$^{4+}$
($t_{2g}^{\,3}e_{g}^0$) and Mn$^{3+}$ ($t_{2g}^{\,3}e_{g}^1$). In
view of the much lower value of $M_S$ observed in the Ca-doped
sample this very simplified picture of considering spin-only
values remains, however, questionable.

\section{Summary}

We have presented a comparative study of the structure,
magnetization and resistivity of charge-carrier doped \lmco\ with
M\,=\,Ca, Sr, and Ba ($0\le x\le 0.3$). The Sr- and Ba-doped
samples crystallize in a rhombohedral structure, which slowly
(moderately fast) approaches cubic symmetry with increasing Sr
(Ba) content. For \lcco\ there is a structural phase transition
from rhombohedral to orthorhombic symmetry for $x\ge 0.2$. For all
three series the resistivity rapidly decreases with increasing
concentration of M. For Ca doping the $\rho(T)$ curves do not
reach a metallic temperature dependence, whereas the Sr-doped
crystals become metallic for $x>0.18$. The Ba-doped samples are
metallic above about 100~K for $x\ge 0.25$, but there are
pronounced localization effects for low temperatures. The
different resistivities may be explained by an increasing
bandwidth with increasing ionic radii of the M$^{2+}$ ions on the
one hand side and by local disorder due to the different ionic
radii of La$^{3+}$ and M$^{2+}$ on the other. The $M(T)$ curves
indicate spin-glass behavior at low temperatures for the Ba- and
Sr-doped samples with $x<0.2$ and $0.18$, respectively, whereas
for larger $x$ ferromagnetic order occurs. In contrast, the
\lcco\ samples show ferromagnetic order over the entire doping
range. The reason for the absence of the spin glass region in
\lcco\ remains to be clarified. The ferromagnetic order occurs at
the largest ordering temperatures $T_c(x)$ for Sr doping, for Ba
doping $T_c(x)$ is slightly smaller and for Ca doping
significantly smaller. Similar to the case of doped
manganites~\cite{rodriguezmartinez96a} this non-monotonous
dependence of $T_c$ on the ionic radii of M$^{2+}$ can be
phenomenologically explained  by assuming that the largest
$T_c(x)$ would be obtained in a perfectly cubic perovskite without
disorder and is reduced by both, a deviation from cubic symmetry
and local disorder due to different radii of La$^{3+}$ and
M$^{2+}$. \lsco\ has higher $T_c$ values, since $T_c$ is strongly
suppressed by a large deviation from cubic symmetry in \lcco\ and
by pronounced local disorder in \lbco .

\begin{acknowledgments}
We acknowledge fruitful discussions with M.\,Braden,
J.\,B.\,Goodenough, M.\,Gr\"{u}ninger, D.\,Khomskii, and
L.\,H.\,Tjeng. We thank G.\,Dhalenne, P.\,Reutler, and
A.\,Revcolevschi for their help when we were growing our first
single crystals. This work was supported by the Deutsche
Forschungsgemeinschaft through SFB~608.
\end{acknowledgments}


\begin{thebibliography}{10}
\bibitem{remark_lcco_sc}
In {R}ef.~[\onlinecite{baily02a}] a
{L}a$_{0.85}${C}ar$_{0.15}${C}o{O}$_{3}$
  single crystal was used, but unfortuantely no details of the single crystal
  growth are given.
\bibitem{remark12fold}
We could also use the {A}-site ionic radii for 12-fold
coordination as is done
  e.g. in {R}ef.~[\onlinecite{dabrowski98a}]. {T}his would cause quantiatively
  different parameters, but the main qualitative conclusions remain unchanged.
\bibitem{remarkitoh}
Since we have not grown a single crystal of
  {L}a$_{0.8}${S}r$_{0.2}${C}o{O}$_{3}$ we used the corresponding
  ${T}_c=180$\,{K} from {R}ef.~[\onlinecite{itoh94a}].
\end{thebibliography}

\end{document}